\documentclass[12pt]{article}
\newcommand{\newc}{\newcommand}
\newc{\ra}{\rightarrow}
\newc{\lra}{\leftrightarrow}
\newc{\beq}{\begin{equation}}
\newc{\eeq}{\end{equation}}
\newc{\barr}{\begin{eqnarray}}
\newc{\earr}{\end{eqnarray}}

\def\al{\alpha}

\def\ket#1{|{#1}\rangle}

\def\braket#1#2{\langle{#1}|{#2}\rangle}
\def\Braket#1#2#3{\langle{#1}|{#2}|{#3}\rangle}

\begin{document}
\begin{titlepage}
\begin{center}

{\large \bf  Closed expressions for the nuclear moments in
semi-leptonic processes}

\vspace{12mm}

V.Ch. CHASIOTI AND T.S. KOSMAS
\vspace{2mm}

{\it Theoretical Physics Division, University of Ioannina,
GR-45110 Ioannina, Greece }
\end{center}
\vspace{7mm}
\begin{abstract}
We discuss a general formalism needed for a unified description of
the weak and electromagnetic processes in nuclei which is based on
the multipole decomposition of the hadronic currents. The use of
harmonic oscillator single-particle basis, which is commonly
employed in many-body nuclear calculations, simplifies the
relevant expressions. We present analytic formulas for the
corresponding radial integrals which enter the transition matrix
elements of one-body operators in various semi-leptonic nuclear
processes. As an example, we apply our formalism in order to
simplify previous formalism giving the nuclear moments for the
$0\nu\beta\beta$ - decay. Our results for the radial integrals
refer to the 1s-0d-1p-0f model space.
\end{abstract}

PACS number(s): 23.20.Js, 23.40.-s, 25.30.-c, 24.10.-i.
\vspace{0.5cm}


\end{titlepage}

\section{Introduction}

In a unified treatment of the semi-leptonic electroweak
interactions in nuclei, like lepton-nucleus scattering
\cite{CNP,DonPe}, $\beta$-decay modes \cite{Suh-Civ},
charged-lepton nuclear capture ($\mu^-$ capture, etc.)
\cite{Suh-Lhe}, neutrino-nucleus induced reactions
\cite{DonPe,CoDoW,Don-Wal}, exotic semi-leptonic processes in
nuclei (double beta-decay \cite{Suh-Fes,Barb98,Barb99}, $\mu^- \ra
e^-$ conversion, etc. \cite{Kos01,KRF}), at first, one takes
advantage of the well known electromagnetic interactions to probe
the nucleus and determine accurately the nuclear charge,
convection current and spin magnetization distributions
\cite{CNP,DonPe,Don-Wal,KV92}. Afterwards, having the nuclear
structure uncertainties reduced to a minimum, one can predict the
cross sections for a variety of semi-leptonic weak interaction
processes.

As is well known, the reaction rates $\Gamma_{i \to f}$ between
the initial $\ket{i}$ and a final $\ket{f}$ discrete nuclear
states of any of the aforementioned processes is written in terms
of the matrix elements of a specific effective Hamiltonians
$H_{eff}$ as
$$\Gamma_{i \to f} \sim \big| \Braket{f}{H_{eff}}{i}\big |^2$$
where the $H_{eff}$ can be assumed to be constructed
within the current-current interaction hypothesis by first writing
down the leptonic ($j_{\lambda}$) and hadronic ($J_{\lambda}$)
currents.

The effective hadronic current-density operator ${\hat {\cal
J}}_\lambda$ which enters the $H_{eff}$, contains only strong
isoscalar ($T=0$) and isovector ($T=1$) pieces and, in general, it
has both vector (V) and axial vector (A) pieces as
\begin{equation}
[{\hat {\cal J}}_\lambda ]^{T M_T} \, =\,
\beta_V^{(T)} [{\hat J}_\lambda  ]^{T M_T} \, +
\, \beta_A^{(T)} [ {\hat J}_\lambda^5  ]^{T M_T}.
\label{projM-M}
\end{equation}
where $\beta_V^{(T)}$ and $\beta_A^{(T)}$ are polar-vector and
axial-vector coefficients involving the relevant couplings (the
purely electromagnetic interactions involve only vector current
density, but the weak interactions involve in addition axial
vector current). In neutral-current processes $M_T=0$ for both
isoscalar and isovector parts, while in charge changing
interactions $T=1$ with $M_T=\pm 1$. One, usually, assumes that
for a nuclear target (denoted as $(A,Z)$ with Z, A the proton- and
mass-number respectively) the states $\ket{i}$ and $\ket{f}$ are
characterized by definite angular momentum and parity
($\ket{\alpha} \equiv \ket{J^\pi}$). In neutral-current reactions
(lepton scattering, $\mu-e$ conversion, etc.) the final state
$|f\rangle$ is, in general, an excited state of the nucleus
$(A,Z)$ while in charged-current processes $\ket{f}$ denotes an
excited state of the $(A,Z\mp 1)$ nucleus.

The nuclear calculations of $\Braket{f}{H_{eff}}{i}$ proceed
usually by decomposing the Fourier transform of the hadronic
current-density operator into irreducible tensors in strong
isospin space \cite{DonPe,Suh-Civ}. Then, spherical Bessel
functions $j_l(x)$ in central or non-central interaction
components are obtained. The nuclear transition rates are
subsequently determined by the matrix elements of these operators.
The possibility to obtain closed expressions for the matrix
elements of the basic one-body operators describing the above
mentioned reactions has been examined in previous theoretical
studies with special emphasis on the use of harmonic oscillator
wave functions \cite{Suh-Fes,Barb98,Barb99,KRF,KV92,KV97}.

In Refs. \cite{Kos01}-\cite{KV97} a method was developed which
provides explicit compact expressions for some basic nuclear form
factors (those needed for the $\mu^-\to e^-$ conversion in nuclei,
scattering of cold dark matter candidates off nuclei, etc.). The
advantage of this method is that it allows the easy calculation of
the required reduced matrix elements by separating the geometrical
coefficients from the kinematical parameters of the reaction. In
this way, the transition matrix elements for every value of the
momentum transfer q can be evaluated immediately. It is worth
remarking that such a compact formalism provides a very useful
insight to those authors who wish to adopt a phenomenological
approach and fit the nuclear transition strengths.

In the present work, we extend the method of Ref.
\cite{Kos01}-\cite{KV97} so as to provide expressions for the
reduced matrix elements of any of the seven basic single-particle
tensor operators describing the above-mentioned semi-leptonic
processes \cite{DonPe,CoDoW,Don-Wal}. We especially deal with the
operators involving the differential operator $\nabla$ (nabla),
which have not been studied previously \cite{KRF,KV92,KV97}. To
this end, we first treat the corresponding radial matrix elements
which contain derivatives with respect to the coordinate r.
Finally, we discuss some representative examples as applications
of our formalism focusing on the improvement of the previous
expressions giving the radial part of the nuclear moments in the
neutrinoless double-beta decay \cite{Suh-Fes}-\cite{Barb99}.

\section{ Multipole decomposition of the hadronic currents}

From a nuclear physics point of view, the hadronic current
($J_{\mu}$) which enters the weak and electromagnetic interaction
Hamiltonian $H_{eff}$ is of primary interest. In our convention
the four-current operator $\hat J_{\mu}$ is written as
\beq
\hat J_{\mu}({\bf r})= (\hat {\rho}({\bf r}), \hat {\bf J}({\bf r}))
\eeq
where $\hat {\rho}({\bf r})$ the density and $\hat {\bf J}({\bf
r})$ the three-current operators. The standard multipole expansion
procedure (see Appendix A) \cite{DonPe,Don-Wal}, applied on the
matrix elements of $\hat J_{\mu}({\bf r})$, leads to spherical
tensor operators which are given in terms of projection functions
involving spherical Bessel functions $j_L(r)$ and spherical
Harmonics $Y^L_M({\hat r})$ or vector spherical Harmonics ${\bf
Y}^{(L,1)J}_M({\hat r})$ as:
\beq
M^J_M(q{\bf r}) \,=\,\delta_{LJ} j_L(qr)Y^L_M(\hat r),
\label{proj1M-M}
\eeq
\beq  {\bf M}_M^{(L1)J}(q{\bf r}) \, =\, j_L(qr){{\bf
Y}}_M^{(L1)J}(\hat r), \label{proj2M-M}
\eeq
where
\beq  {\bf Y}_M^{(L1)J}(\hat r) \, = \, \sum_{m,q} \braket
{Lm1q}{JM} {Y^L_m}(\hat r)  {\hat e}_q, \label{proj-vsh} \eeq
${\hat e}_q$ is a unit vector in the direction of the
three-momentum transfer {\bf q}. The magnitude of $\bf q$ ($q=
\vert \bf q\vert$) is given from the kinematics of the studied
process (see e.g. Ref. \cite{Kos01}).

Using the projections functions Eqs. (\ref{proj1M-M}) and
(\ref{proj2M-M}), in the case of the polar-vector current (${\hat
J}_\lambda$) we obtain the multipole operators
\beq
\hat M_{JM;TM_T}^{coul} \, = \, \int d{\bf r} M^J_M(q{\bf r})
\hat {\rho}({\bf r})_{TM_T},
\label{Coul-op}
\eeq
\beq
\hat L_{JM;TM_T} \, = \, i\int d{\bf r}\left(\frac{1}{q}
\nabla  M^J_M(q{\bf r})\right) \cdot \hat {\bf J}({\bf r})_{TM_T},
\label{Long-op}
\eeq
\beq
\hat T_{JM;TM_T}^{el} \, = \, \int d{\bf r}\left(\frac{1}{q}
\nabla \times {\bf M}^{JJ}_M(q{\bf r})\right)
\cdot \hat {\bf J}({\bf r})_{TM_T},
\label{Tr-el-op}
\eeq
\beq
\hat T_{JM;TM_T}^{mag} \, = \, \int d{\bf r} {\bf M}^{JJ}_M(q{\bf r})
\cdot \hat {\bf J}({\bf r})_{TM_T}.
\label{Tr-ma-op}
\eeq
which have well-defined parity and are called as: Coulomb,
longitudinal, transverse-electric, and transverse-magnetic
multipoles, respectively. The first three operators have parity
$(-)^J$ (normal parity operators), while the parity of $\hat
T_J^{mag}$ is $(-)^{J+1}$ (abnormal parity operator). The
analogous axial-vector multipoles are
\beq
\hat M_{JM;TM_T}^5 \, = \, \int d{\bf r} M^J_M(q{\bf r})
{\hat \rho}({\bf r})^5_{TM_T},
\label{Coul5-op}
\eeq
\beq
\hat L_{JM;TM_T}^5 \, = \, i\int d{\bf r}\left(\frac{1}{q}
\nabla  M^J_M(q{\bf r})\right) \cdot \hat {\bf J}({\bf r})^5_{TM_T},
\label{Long5-op}
\eeq
\beq
\hat T_{JM;TM_T}^{el5} \, = \, \int d{\bf r}\left(\frac{1}{q}
\nabla \times {\bf M}^{JJ}_M(q{\bf r})\right)
\cdot \hat {\bf J}({\bf r})^5_{TM_T},
\label{Tr-el5-op}
\eeq
\beq
\hat T_{JM;TM_T}^{mag5} \, = \, \int d{\bf r} {\bf M}^{JJ}_M(q{\bf r})
\cdot \hat {\bf J}({\bf r})^5_{TM_T}.
\label{Tr-ma5-op}
\eeq
The first three axial-vector multipoles have parity $(-)^{J+1}$
while $\hat T_J^{mag5}$ is a normal parity operator. For a
conserved vector current (CVC) like the electromagnetic, the
longitudinal multipoles can be written in terms of the Coulomb
ones as
\beq \hat
L_{JM_J;Tm_T}(q)= \frac{q_0}{q} \hat M_{JM_J;TM_T}(q) \eeq
($q_0$ is the time component of the four-momentum transfer, $q_\mu
= (q_0, {\bf q})$). In this case, the number of independent
operators resulting from the projection procedure is reduced to
seven.

The isospin dependence of the operators
(\ref{Coul-op})-(\ref{Tr-ma5-op}) includes the operator
$I_T^{M_T}$ given by \cite{DonPe}
\beq I_T^{M_T} \,=\, \left\{\begin{array} {l
@{\quad} l} 1, \qquad \qquad\qquad\qquad \,\,
\,\,T=0, \,M_T=0 \\
\tau_0 = \tau_3, \qquad \qquad \qquad\, T=1, \,M_T=0 \\
\tau_{\pm}= \mp \frac{1}{\sqrt{2}}(\tau_1 \pm \tau_2), \quad
T=1, \,M_T= \pm 1 \\
\end{array}\right.
\eeq
The exact form of $I_T^{M_T}$ is determined by the specific
semi-leptonic reaction in question (see the Introduction).

The matrix elements of the seven basic operators of Eqs.
(\ref{Coul-op})-(\ref{Tr-ma5-op}) involve isospin dependent form
factors $F_X^{(T)}$ (see Appendix A). For this reason, we define
seven new operators of which the spin-space parts are written (in
the first quantization) as \cite{DonPe,CoDoW,Don-Wal}:
\barr T_1^{JM} \, \equiv \, M^J_M(q{\bf r}) \, = \, \delta_{LJ}\,
j_L(\rho)Y^L_M(\hat r), \label{opM} \earr \barr T_2^{JM} \,
\equiv\, {\Sigma}^J_{M}(q{\bf r}) \,=\,{{\bf M}}^{JJ}_{M}\cdot
 \mbox{\boldmath $ \sigma $},
\label{opSig}
\earr
\barr
T_3^{JM} \, \equiv \, {\Sigma^{\prime}}^J_{M}(q{\bf r})  \,= \,
-i\left\{\frac{1}{q}
\nabla \times {{\bf M}}^{JJ}_{M}(q{\bf r})\right\} \cdot
 \mbox{\boldmath $ \sigma $},
\label{opSigp}
\earr
\barr
T_4^{JM} \, \equiv \,{\Sigma  ''}^J_{M}(q{\bf r}) \, = \,
 \Big\{\frac{1}{q}\nabla
M^J_{M}(q{\bf r})\Big\} \cdot
\mbox{\boldmath  $ \sigma $},
\label{opSigpp}
\earr
\barr
T_5^{JM} \, \equiv \,{\Delta}^J_{M}(q{\bf r}) \,= \,
{{\bf M}}^{JJ}_{M}(q{\bf r})
\cdot\frac{1}{q}\nabla,
\label{opDel}
\earr
\barr
T_6^{JM} \, \equiv \,{\Delta ^{ \prime}}^J_{M}(q{\bf r}) \,= \,
-i\Big\{\frac{1}{q}\nabla \times {{\bf M}}^{JJ}_{M}(q{\bf r})\Big\}
 \cdot \nabla,
\label{opDelp} \earr \barr T_7^{JM} \, \equiv \,\Omega^J_{M}(q{\bf
r}) \, = \, M^J_{M}(q{\bf r})\mbox{\boldmath  $ \sigma $} \cdot
\frac{1}{q}\nabla.
\label{opOm}
\earr
(hereafter we use the shorter notation for the projection
functions of Eqs. (\ref{proj2M-M}) and (\ref{proj-vsh}) as ${\bf
M}^{LJ}_M$ and ${\bf Y}^{LJ}_M$, respectively). For the reader's
convenience, in Eqs. (\ref{opM})-(\ref{opOm}) we adopt the usual
notation \cite{DonPe}. Using properties of the nabla operator
($\nabla$), Eqs. (\ref{opSigp}), (\ref{opSigpp}) and
(\ref{opDelp}) can be rewritten as:
\beq T_3^{JM} \,
\equiv\,{\Sigma^{\prime}}^J_{M} \, = \, [J]^{-1} \Big\{-J^{1/2}
{\bf M}^{J+1J}_{M} + (J+1)^{1/2} {\bf M}^{J-1J}_{M}\Big\} \cdot
 \mbox{\boldmath $ \sigma $},
\label{opSigpt} \eeq \beq T_4^{JM}
\,\equiv\,{\Sigma^{\prime\prime}}^J_M \, = \,[J]^{-1}
\Big\{(J+1)^{1/2}{{\bf M}}^{J+1J}_{M} + J^{1/2}{{\bf
M}}^{J-1J}_{M}\Big\} \cdot \mbox{\boldmath  $ \sigma $},
\label{opSigppt} \eeq
\beq
T_6^{JM} \, \equiv\,{\Delta ^{ \prime}}^J_{M} \,= \,[J]^{-1}
\Big\{-J^{1/2}{{\bf M}}^{J+1J}_{M} + (J+1)^{1/2}{{\bf
M}}^{J-1J}_{M}\Big\} \cdot \frac{1}{q}\nabla,
\label{opDelpt}
\eeq
(throughout this work we use the common symbol $[J] \, =
\,(2J+1)^{1/2}$). By glancing at Eqs. (\ref{opM})-(\ref{opDelpt}),
we conclude that, the seven basic single-particle operators of
definite parity are built with the aid of four operators
\beq {\mathcal O}_1^{JM} \, = \, M^J_M,
\quad {\mathcal O}_2^{JM} \, = \,{\bf
M}^{LJ}_M\cdot\mbox{\boldmath $\sigma$}, \label{O_1,2M} \eeq \beq
{\mathcal O}_3^{JM} \, =\, {\bf M}^{LJ}_M\cdot\frac{1}{q}\nabla,
\quad {\mathcal O}_4^{JM} \, = \, M^J_M\mbox{\boldmath $\sigma$}
\cdot\frac{1}{q}\nabla. \label{O_3,4Mdel} \eeq
which contain the projection functions ${M}^{J}_M$ and ${\bf
M}^{LJ}_M$ or their products with the nucleon-spin {\boldmath
$\sigma$} and/or the $\nabla$ operator.

\section{The Single-particle matrix elements }

Many quantities of the semi-leptonic electroweak processes are
expressed (to a good approximation) in terms of single-particle
nuclear matrix elements of the one-body operators $T_i^{JM},
i=1,2,...7$ \cite{DonPe,Suh-Civ,CoDoW,Don-Wal}. These matrix
elements, by applying the Wigner-Eckart theorem, are written as:
\beq \Braket{j_1m_1}{T^{JM}_i}{j_2m_2}  \, = \,(-)^{j_1-m_1}
\pmatrix{j_1  &  J  &  j_2 \cr
         -m_1 & M & m_2 \cr}
\Braket{j_1}{|T^J_i|}{j_2}, \label{Wig-Eck} \eeq
In our notation, the single-particle wave functions are labelled
(see Appendix B) with the quantum numbers ($n l s j m_{j}$). The
reduced matrix elements of the operators ${\mathcal O}_i^J$,
$i=1,2,3,4$, after applying the re-coupling relations, can be
written in the closed forms shown below.

1. For the operators ${\mathcal O}_i^J$, with $i=1,2$, the reduced
matrix elements $\Braket{j_1}{|{\mathcal O}_i^{(L,S_i)J}|}{j_2}$
have been compactly written as \cite{KRF,KV97}
\beq
\Braket{j_1}{|{\mathcal O}_i^{(L,S_i)J}|}{j_2}\,= \,
(l_1 \  L \  l_2)\,{\mathcal U}^J_{LS}
 \Braket{n_1l_1}{j_L(\rho)}{n_2l_2}, \quad i=1,2
\label{gen-ME}
\eeq
where the symbols $(l_1 \ L \ l_2)$ and ${\mathcal U}_{LS}^J$ are
\beq (l_1 \ L \ l_2) \, \equiv \, (-)^{l_1}\frac{1}{\sqrt{4\pi}}
[l_1][L][l_2] \pmatrix{l_1 & L & l_2 \cr
                        0  & 0 &  0 \cr },
\label{3-jcoef}
\eeq
\beq
{\mathcal U}_{LS}^J \, \equiv \,
 [j_1][j_2][J](S+1)^{1/2}(S+2)^{1/2}
 \left\{\matrix{l_1 & l_2 & L \cr
                  1/2 & 1/2 & S \cr
                  j_1 & j_2 & J \cr}\right\}.
\label{9-jcoef} \eeq
In the case of the Fermi-type operator ${\mathcal O}_1^J$, we must
put $S=0$ in the 9-j symbol of Eq. (\ref{9-jcoef}) while in the
case of the Gamow-Teller operator ${\mathcal O}_2^J$, $S=1$.

2. The reduced matrix elements of ${\mathcal O}_3^J$ after some
manipulation can be cast in the form
\beq
\Braket{j_1}{|{\bf M}^{LJ}(q{\bf r}) \cdot \frac{1}{q}
\nabla|}{j_2} = \sum_{\al} {\mathcal A}_L^{\al}(j_1 j_2 ;J)
\Braket{n_1l_1}{\theta _L^{\al}(\rho)}{n_2l_2}, \quad \al = \pm
\label{MdelME}
\eeq
where the coefficients ${\mathcal A}_L^{\pm}$
are
\barr
{\mathcal A}_L^{\pm}(j_1 j_2; J) \, & = &
\,\pm(-)^{l_1+L+j_2+1/2} [j_1][j_2][J] \left(\frac{2l_2+1 \mp
1}{2}\right)^{1/2} (l_1 \  L \  l_2 \mp 1)
\nonumber \\
&& \times {\mathcal W}_6(l_1, j_1, 1/2, j_2, l_2, J) {\mathcal
W}_6( L, 1, J, l_2, l_1, l_2 \mp 1), \label{CfMdelME} \earr with
${\mathcal W}_6$ representing the common 6-j symbol \beq {\mathcal
W}_6(l_1, j_1, 1/2, j_2, l_2, J) \, \equiv \, \left\{ \matrix{l_1
& j_1 & 1/2 \cr
               j_2 & l_2 & J \cr}\right\}.
\label{6-j}
\eeq

3. Similarly, for the  reduced matrix element of ${\mathcal
O}_4^J$ we can write
\beq \Braket{j_1}{|M^J(q{\bf r})\mbox{\boldmath
$\sigma$}
 \cdot \frac{1}{q} \nabla|}{j_2} \,= \,
\sum_{\al} {\mathcal B}_L^{\al}(j_1 j_2; J)
\Braket{n_1l_1}{\theta _J^{\al}(\rho)}{n_2l_2}, \quad \al = \pm
\label{MspdelME}
\eeq
where
\barr
{\mathcal B}_L^{\pm}(j_1 j_2; J) \,& = &\,\pm
\,{\delta}_{j_2,l_2\mp{1/2}}\,
[j_1][j_2] (l_1 \  J \  2j_2 -l_2) \,
{\mathcal W}_6(l_1, j_1, 1/2, j_2, 2j_2-l_2, J).
\label{CfMspdelME}
\earr

From Eqs. (\ref{gen-ME}), (\ref{MdelME}) and (\ref{MspdelME}) we
notice that all basic single-particle reduced matrix elements
required for our purposes rely on the following three types of
radial integrals:
\beq
\Braket{n_1l_1}{\theta^{\al}_l(\rho)}{n_2l_2} \,\equiv \,\int dr
r^2 R_{n_1l_1}^{*}(r)\theta^{\al}_l(\rho)R_{n_2l_2}(r),
\hspace{1.5cm} \al \, = \,0,\pm \label{rad-int} \eeq with \beq
\theta^0_l(\rho) = j_l(\rho),\qquad \theta^{\pm}_l(\rho)  =
j_l(\rho)\left(\frac{d}{d\rho} \pm \frac{2l_2+1 \pm
1}{2\rho}\right). \label{theta} \eeq
The argument $\rho$ in Eqs. (\ref{theta}) is equal to $\rho = qr$.
We note that the derivatives with respect to $\rho$ appeared in
the radial matrix elements come from the application of the
gradient formula in the matrix elements of Eqs. (\ref{MdelME}) and
(\ref{MspdelME}).

\subsection{Expressions for the Radial Integrals}

It is well known that, for single-particle wave functions with
arbitrary radial dependence it is not easy to perform analytically
the integrations over r in Eq. (\ref{rad-int}). These integrals,
however, can be simplified in the case when harmonic oscillator
basis (see Appendix B) is used \cite{DonPe,KV97}. Then they take
the elegant expressions shown below.

(i) For the operator $\theta_l^0(\rho)$ in Ref. \cite{KV97} it was
proved that
\beq
\Braket{n_1l_1}{j_L(\rho)}{n_2l_2} \, = \,
e^{-y}y^{L/2} \sum_{\mu=0}^{n_{max}}\varepsilon_{\mu}^L \,y^{\mu}
, \qquad y=(qb/2)^2
\label{rad-int0}
\eeq
$$ n_{max}=(N_1+N_2-L)/2,$$
where $N_i=2n_i+l_i$ represent the harmonic oscillator quanta of
the $i_{th}$ level. The coefficients
$\varepsilon_{\mu}^L(n_1l_1n_2l_2)$ are given by
\beq \varepsilon_{\mu}^L(n_1l_1n_2l_2)\,=\,G\frac{{\pi}^{1/2}}{2}
\sum_{m_1=\phi}^{n_1}\sum_{m_2=\sigma}^{n_2} n!
\Lambda_{m_1}(n_1l_1)\Lambda_{m_2}(n_2l_2)\Lambda_{\mu}(nL),
\label{cf-eps} \eeq
with
$$ n=m_1+m_2+(l_1+l_2-L)/2. $$
Also, $G \,=\, b^3 N_{n_1l_1}N_{n_2l_2}/2$, where $N_{n_1l_1}$ is
defined in Appendix B. The other symbols of Eq. (\ref{cf-eps}) are
explained in Ref. \cite{KV97}.

(ii) The formulation of the radial matrix elements which include
differential operators $\theta_l^{\pm}$, which constitute the main
task of our present paper, proceeds in a similar manner to that of
Eq. (\ref{rad-int0}). This leads to the expressions
\beq
\Braket{n_1l_1}{j_L(\rho)\left(\frac{d}{d\rho} \pm
\frac{2l_2 + 1 \pm 1}{2\rho}\right)}
{n_2l_2} \, = \, e^{-y}y^{(L-1)/2}
\sum_{\mu=0}^{n_{max}}{\zeta}_{\mu}^{\pm}(L) \,y^{\mu},
\label{rad-int-pm}
\eeq
where the geometrical coefficients
$\zeta_{\mu}^{\pm}(n_1l_1n_2l_2;L)$ are given in terms of those of
Eq. (\ref{cf-eps}), i.e.
\beq {\zeta}_{\mu}^-(L)\,=\, -\frac{1}{2}\left\{\begin{array}{
l@{\quad} l} (n_2+l_2+3/2)^{1/2} \varepsilon_{\mu}^L
(n_1l_1n_2l_2+1)
 + n_2^{1/2}\varepsilon_{\mu}^L (n_1l_1n_2-1l_2+1),&
0\leq \mu< n_{max} \\
(n_2+l_2+3/2)^{1/2}\varepsilon_{n_{max}}^L(n_1l_1n_2l_2+1),  \qquad
\qquad  \mu = n_{max} \\
\end{array} \right.
\label{cf-zetami}
\eeq
\beq {\zeta}_{\mu}^+(L)\,=\, \frac{1}{2} \left\{
\begin{array}{ l@{\quad} l} (n_2+l_2+1/2)^{1/2}
\varepsilon_{\mu}^L(n_1l_1n_2l_2-1)
+  (n_2+1)^{1/2}\varepsilon_{\mu}^L(n_1l_1n_2+1l_2-1),  &
0\leq \mu <n_{max} \\
(n_2+1)^{1/2}\varepsilon_{n_{max}}^L(n_1l_1n_2+1l_2-1),
\qquad\qquad \mu = n_{max}\\
\end{array} \right. \label{cf-etami}
\eeq
The value of the index $n_{max}$ in the latter two cases of radial
integrals is determined by
$$ n_{max}=(N_1+N_2-L+1)/2. $$

At this point, it should be noted that, the explicit and general
formulas of Eqs. (\ref{cf-eps}), (\ref{cf-zetami}) and
(\ref{cf-etami}) hold for every combination of the levels
$(n_1l_1)j_1$, $(n_2l_2)j_2$, and give the geometrical (momentum
independent) coefficients of the polynomials defined in Eqs.
(\ref{rad-int0}) and (\ref{rad-int-pm}). These closed expressions
have been constructed by inverting properly the multiple
summations involved in the corresponding matrix elements (see Ref.
\cite{HaKo02a,HaKo02b}), so as the final summation is performed
over the harmonic oscillator quantum number $N=2n+l$.

In order to show the advantages of the above formalism we discuss
below some applications.

\section{Applications}

As mentioned before, our formalism is applicable for an infinite
single-particle basis and can be applied to any semi-leptonic
electroweak process which take place in the field of nuclei
\cite{HaKo02a,HaKo02b}. In a realistic case one is forced to
truncate the basis set. Thus, e.g. the description of the nuclear
moments for the neutrinoless double beta decay $^{48}Ca \to
^{48}$Ti can be done using a model space including the seven
orbitals of the major shells 2$\hbar \omega$ and 3$\hbar \omega$
\cite{Barb98,Barb99}.

In Tables 1-3 we list the coefficients $\varepsilon _{\mu}^L$ and
$\zeta_{\mu}^{\pm}$ needed to evaluate the radial matrix elements
$\Braket{n_1l_1}{\theta_L^\alpha(\rho)}{n_2l_2}$ in the above
(1s-0d-1p-0f) model space. For some additional geometrical
coefficients $\varepsilon _{\mu}^L$, which determine the
transition matrix elements for scattering of dark matter
candidates off nuclei in the model space including the orbitals
2p-0h and 0i-1g, the reader is referred to Ref. \cite{6HelSy}.

By exploiting the analytic expressions of the Secs. 2 and 3, in
the next subsection we simplify the formalism of Ref.
\cite{Barb98,Barb99} constructed for the calculation of the
nuclear moments in neutrinoless double-beta decay.

\subsection{Nuclear moments for the neutrinoless double-beta decay}

In the description of the nuclear moments for the neutrinoless
double-beta decay \cite{Suh-Fes}-\cite{Barb98} the following
radial integrals are needed:
\beq {\cal
R}^{\kappa}_L(n_1l_1n_2l_2;q) \, =\,
\Braket{n_1l_1}{j_L(qr)r^{\kappa}}{n_2l_2} \eeq
\beq {\cal
R}^{\kappa}_{L_1L_2}(n_1l_1n_2l_2,n_1^{\prime}l_1^{\prime}
n_2^{\prime}l_2^{\prime};\omega) \, =\, \int  q^{2+\kappa}
\upsilon(q;\omega) {\cal R}_{L_1}^0(n_1l_1n_2l_2;q)
 {\cal R}_{L_2}^0(n_1^{\prime}l_1^{\prime}n_2^{\prime}l_2^{\prime};q) dq
\label{2bodyME} \eeq
where $\upsilon(q;\omega)$ are functions describing the energy of
the intermediate neutrinos in various gauge models (light-neutrino
mixing, etc.) \cite{Suh-Civ,Suh-Fes,Barb98,Barb99}. The parameter
$\omega$ is related to the excitation energy of the intermediate
nucleus. Using the compact formalism presented in the previous
section the above integrals are simplified and written as follows:

{\bf a.} For the radial moments ${\cal
R}^{\kappa}_L(n_1l_1n_2l_2;q) $ working as in Eq. (\ref{rad-int0})
we find
\barr {\cal R}^{\kappa}_L(n_1l_1n_2l_2;q) \,=\,
b^{\kappa}y^{L/2}e^{-y}\sum_{\mu=0}^{n_{max}} \varepsilon_{\mu}^L
y^{\mu} \label{rad-mom1} \earr
which implies that the nuclear moments ${\cal
R}_L^{\kappa}(n_1l_1n_2l_2;q)$ are simply obtained with the aid of
the coefficients $\varepsilon_{\mu}^L(n_1l_1n_2l_2)$ of Eq.
(\ref{cf-eps}). Note that for $\kappa = 0$ Eq. (\ref{rad-mom1})
reduces to Eq. (\ref{rad-int0}).

{\bf b.} By replacing the radial integrals in the right-hand side
of Eq. (\ref{2bodyME}) with Eq. (\ref{rad-int0}) and manipulating
their product we take
\barr {\cal R}^0_{L_1}(n_1l_1n_2l_2;q) 
{\cal R}^0_{L_2}(n_1^{\prime}l_1^{\prime}n_2^{\prime}l_2^{\prime};q) \, =\,
e^{-2y} y^{(L_1+L_2)/2}\sum_{\mu=0}^{n_{max}}c_{\mu}y^{\mu}
\label{2body-radm} \earr
where
$$
n_{max}=n_{1max}+n_{2max}, $$ with
$$
n_{1max}=n_1+n_2+(l_1+l_2-L_1)/2 \, , \qquad
n_{2max}=n_1^{\prime}+n_2^{\prime}
+(l_1^{\prime}+l_2^{\prime}-L_2)/2
$$
The coefficients $c_\mu$ in Eq. (\ref{2body-radm}) are obtained
from those of $\varepsilon_{\nu}^{L}$ as
\begin{equation}
c_{\mu}\,=\, \sum_{\lambda=0}^{\mu}
\varepsilon_{\nu}^{L_1}\varepsilon_{\mu -\nu}^{L_2} \label{ci-mu}
\end{equation}
In the last summation $\varepsilon_{\nu}^{L_1} \,= \,0$,  for
$n_{1max}<\nu \leq n_{max}$, and $\varepsilon_{\mu-\nu}^{L_2} \,
=\,0$, for $ n_{2max}<\mu - \nu \leq n_{max}$. By inserting Eq.
(\ref{2body-radm}) into Eq. (\ref{2bodyME}) and putting
$q\,=\,2y^{1/2}/b$ [see Eq. (\ref{rad-int0})], we have
\begin{equation}
{\cal R}^{\kappa}_{L_1L_2}(n_1l_1n_2l_2,n^{\prime}_1 l^{\prime}_1
n^{\prime}_2 l^{\prime}_2;\omega)\, = \,
\frac{2^{\kappa+2}}{b^{\kappa+3}} \sum_{\mu=0}^{n_{max}}c_{\mu} \,
{\cal I}^{\lambda}(\omega) \label{2b-radm}
\end{equation}
where  $$\lambda = \mu + (L_1 + L_2 +\kappa +1)/2$$ The quantities
${\cal I}^{\lambda}(\omega)$ represent the integrals
\begin{equation}
{\cal I}^\lambda(\omega) \, = \, \int_0^{\infty}
\upsilon(y;\omega) e^{-2y}y^\lambda dy. \label{int2b}
\end{equation}
which can be computed after providing the exact form of
$\upsilon(y;\omega)$ \cite{Suh-Civ}. Equation (\ref{2b-radm}) is
much simpler than Eq. (D.1) of Ref. \cite{Barb98}, since  the
latter does not include a simple summation as Eq. (\ref{2b-radm}),
but a double one over similar integrals to those of Eq.
(\ref{int2b}).

\section{Summary and Conclusions}

Our main concern in the present paper was the general formalism
needed for calculating the weak and electromagnetic processes in
nuclei. Starting from the decomposition of the hadronic current
into tensor multipole operators, we investigated the possibility
of constructing analytic formulas for the matrix elements of the
principal semi-leptonic operators. We payed special attention on
the multipole matrix elements taken between single-particle states
and restricted ourselves on the formulation of their radial part.

By utilizing harmonic oscillator basis, we have achieved all types
of radial moments related to the seven basic one-body operators
(involved in semi-leptonic reactions), to be written in the form
of closed analytic expressions, i.e. as products of an exponential
times a polynomial of even powers in the momentum transfer q. The
coefficients of the polynomials found are in general simple
numbers (for diagonal matrix elements they are always rational
numbers) and they can be calculated with simple codes. We applied
the above compact expressions in order to simplify the previous
formalism \cite{Suh-Fes,Barb98,Barb99} giving the nuclear moments
in the interesting neutrinoless double-beta decay process. This
helped us to see the advantages of the formalism described in
Secs. 3 and 4.

\section{Acknowledgements}

This work has been supported in part by the Research Committee of
the University of Ioannina.

\begin{center}
{\bf Appendix A}
\end{center}
{\bf 1.} The identities which are needed in the multipole
decomposition procedure are
\beq
exp(-i {\bf q.r}) \,=\,
4\pi\sum_{JM_J}(-i)^JM_M^J({\bf r}) Y^*_{JM}({\hat e}_{q_0})
\eeq
\beq {\hat e}_{q_{\lambda}}exp(-i{\bf q.r}) \, \equiv \,
-(2\pi)^{1/2} \sum_{J\geq 1}(-i)^J(2J+1)^{1/2}\Big[\lambda {\bf
M}_M^{JJ}({\bf r}) + \frac{1}{q} \nabla \times {\bf M}_M^{JJ}({\bf
r})\Big] \eeq
where $\lambda = \pm $ and the unit vector
${\hat e}_{q_{\lambda}}$ has the spherical components
\beq {\hat e}_{q_0}\,\equiv\, {\hat e}_z\,=\, {\bf q}/q
\eeq
\beq
{\hat e}_{q_{\pm}} \, \equiv\, \mp
\frac{1}{\sqrt{2}}({\hat e}_x \pm i{\hat e}_y)
\eeq
The components of the rank-one spherical tensor $\nabla_{\mu}$ are
\beq \nabla_0\,=\,\frac{\partial}{\partial z}
\eeq
\beq
\nabla_{\pm}\,=\, \mp \left(\frac{\partial}{\partial x} \pm
i\frac{\partial}{\partial y}\right)
\eeq

{\bf 2.} For a single free nucleon  (using Lorentz covariance,
conservation of
parity, time-reversal invariance, and isospin invariance) the matrix
elements  of the vector and axial-vector currents, in spin-isospin
space, are written as
\barr
\Braket{{\bf k}'\lambda ';1/2m_{t'}}{J_{\mu}(0)_{TM_T}}
{{\bf k}\lambda;1/2m_t} \,& = &\, i{\bar u}({\bf k}'\lambda ')
\left[F_1^{(T)}\gamma_{\mu} + F_2^{(T)}\sigma_{\mu\nu}q_{\nu} +
iF_S^{(T)}q_{\mu}\right]u({\bf k} \lambda)
 \nonumber \\
& \times &
\Braket{1/2m_{t'}}{I_T^{M_T}}{1/2m_t}
\earr
\barr
\Braket{{\bf k}'\lambda ';1/2m_{t'}}{J_{\mu}^5(0)_{TM_T}}
{{\bf k}\lambda;1/2m_t} \,&=&\, i{\bar u}({\bf k}'\lambda ')
\left[F_A^{(T)}\gamma_5 \gamma_{\mu} - iF_P^{(T)}\gamma_5q_{\mu} -
F_T^{(T)}\gamma_5\sigma_{\mu\nu}q_{\nu}\right]u({\bf k} \lambda)
\qquad
 \nonumber \\
& \times & \Braket{1/2m_{t'}}{I_T^{M_T}}{1/2m_t} \earr
The plane-wave single-nucleon states are labelled with 3-momenta
${\bf k}({\bf k}')$, helicities $\lambda (\lambda ')$ and isospins
$1/2m_t (1/2m_{t'})$. The single-nucleon form factors $F_X^{(T)}=
F_X^{(T)}(q_{\mu}^2)$, with $T=0,1$, and  $X=1,2,S,A,P,T$ (vector
(Dirac), vector (Pauli), scalar, axial, pseudoscalar,and tensor)
are all functions of the momentum transfer $q_{\mu}^2$.

\begin{center}
{\bf Appendix B}
\end{center}

The radial part of the wave functions in a (three-dimensional
isotropic) harmonic oscillator potential is written as
\barr
 R_{nl} \, = \,N_{nl}x^le^{-x^2/2}{\cal L}_n^{l+1/2}(x^2),
\earr
where $x =r/b$, with $b$ is the harmonic oscillator size parameter
and $N_{nl}$ the normalization factor
\barr N_{nl}^2 \, =
\,\frac{2n!}{b^3\Gamma(n+l+3/2)}. \earr
$\Gamma(x)$ denotes the known gamma function and  ${\cal
L}_n^{l+1/2}$ represent the Laguerre polynomials defined by:
\barr{\cal L}_n^{l+1/2}(x) \,= \,\sum_{m=0}^n \Lambda_m(nl) \,x^m
\,=\, \sum_{m=0}^n \frac{(-)^m}{m!} \pmatrix{n + l +1/2 \cr n -
m\cr}\,x^m \, . \earr

In Sect. 3, by writing $\ket{n(l1/2)jm_{j}}$, we mean
$R_{nlj}(r)[Y_l(\Omega_r)\otimes \chi_{1/2}]_{m_j}^j$ with
$\chi_{1/2}$ the Pauli spinor. We note that for the harmonic
oscillator it holds $R_{nlj}(r) = R_{nl}(r)$. The adopted, in the
present work, sequence of single-particle states is: 0s1/2, 0p3/2,
0p1/2, ...

\begin{table}
\begin{center}
\begin{tabular}{|c|c|c|c|c|c|}  \hline\hline
 & & & & & \\
 {\em $n_1l_1 - n_2l_2$} &
 \multicolumn{1}{c|}{\em L} &
 \multicolumn{1}{c|}{\em $\mu=0$} &
 \multicolumn{1}{c|}{\em ${\mu=1}$}  &
 \multicolumn{1}{c|}{\em ${\mu=2}$}  &
 \multicolumn{1}{c|}{\em ${\mu=3}$}  \\ \hline
 & & & & & \\
$0d - 0d$ & 0 & 1 & -$\frac{4}{3}$ & $\frac{4}{15}$  &
                                                       \\ & & & & & \\
          & 2 & $\frac{14}{15}$ & -$\frac{4}{15}$ &   & \\ & & & & & \\
          & 4 & $\frac{4}{15}$ &  &   & \\ & & & & & \\
$0f - 0f$ & 0 & 1 & -2 & $\frac{4}{5}$  & -$\frac{8}{105}$
                                                       \\ & & & & & \\
          & 2 & $\frac{6}{5}$ & $\frac{24}{35}$ & $\frac{8}{105}$
                                             & \\ & & & & & \\
          & 4 & $\frac{44}{105}$ & -$\frac{8}{105}$  &   & \\ & & & & & \\
          & 6 & $\frac{8}{105}$ &   &   & \\ & & & & & \\
$1s - 1s$ & 0 & 1 & -$\frac{4}{3}$ & $\frac{2}{3}$  &
                                     \\ & & & & & \\
$1p - 1p$ & 0 & 1 & -2 & $\frac{22}{15}$  & -$\frac{4}{15}$ \\ & & & & & \\
       & 2 & $\frac{6}{5}$  &  $\frac{16}{15}$  & -$\frac{4}{15}$ &
                                     \\ & & & & & \\
$0d - 1s$ & 2 & 1 & -$\frac{8}{3}\sqrt{\frac{1}{10}}$ & 
                           $\frac{4}{3}\sqrt{\frac{1}{10}}$  &
                                     \\ & & & & & \\
$0d - 0f$ & 1 & $\frac{1}{3}\sqrt{14}$ & -$\frac{4}{15}\sqrt{14}$ &
                  $\frac{4}{105}\sqrt{14}$  & \\ & & & & & \\
       & 3 & $\frac{6}{35}\sqrt{14}$  &  -$\frac{4}{105}\sqrt{14}$  &  &
                                     \\ & & & & & \\
$0d - 1p$ & 1 & -$\frac{2}{3}$ & $\frac{6}{5}$ &
                 -$\frac{4}{15}$  & \\ & & & & & \\
       & 3 & -$\frac{8}{15}$  &  $\frac{4}{15}$  &  &
                                     \\ & & & & & \\
$1s - 1p$ & 1 & $\frac{\sqrt{10}}{3}$ & -2$\sqrt\frac{2}{5}$
                    & $\frac{3}{2}\sqrt{\frac{2}{5}}$ & \\ & & & & & \\
$0f - 1p$ & 2 & $\frac{4}{15}\sqrt{14}$ & $\frac{26}{105}\sqrt{14}$ &
                 -$\frac{4}{105}\sqrt{14}$  & \\ & & & & & \\
       & 4 & -$\frac{4}{105}\sqrt{14}$  &  $\frac{4}{105}\sqrt{14}$  &  &
                                     \\ & & & & & \\ \hline
\end{tabular}
\caption{Coefficients $\varepsilon_{\mu}^L$ which determine the
radial nuclear moments $\Braket{n_1l_1}{j_L(q r)
r^\kappa}{n_2l_2}$ (see Eqs. \protect (\ref{rad-int0}) and
\protect (\ref{rad-mom1})) in the 1s-0d and 1p-0f model space. }
 \end{center}
\end{table}
\begin{table}
\begin{center}
\begin{tabular}{|c|c|c|c|c|c|}  \hline\hline
& & & & & \\
{\em $n_1l_1 - n_2l_2$} & \multicolumn{1}{c|}{\em L} &
 \multicolumn{1}{c|}{\em $\mu=0$} &
 \multicolumn{1}{c|}{\em ${\mu=1}$}  &
 \multicolumn{1}{c|}{\em ${\mu=2}$}  &
 \multicolumn{1}{c|}{\em ${\mu=3}$}  \\ \hline
 & & & & & \\
$0d - 0d$ &  1 & -$\frac{7}{6}$ & $\frac{14}{15}$ & -$\frac{2}{15}$  &
                                                       \\ & & & & & \\
          & 3 & -$\frac{3}{5}$ & $\frac{2}{15}$ &   & \\ & & & & & \\
$0f - 0f$ & 1 & -$\frac{3}{2}$ & $\frac{9}{5}$ &
                            -$\frac{18}{35}$ & $\frac{4}{105}$ \\ & & & & & \\
    & 3 & -$\frac{33}{35}$ & $\frac{44}{105}$  & -$\frac{4}{105}$ &
                                                        \\ & & & & & \\
    & 5 & -$\frac{26}{105}$ & $\frac{4}{105}$  &   &
                                                        \\ & & & & & \\
$1p - 1p$    & 1 & -$\frac{5}{6}$ & $\frac{19}{15}$ &
                                  -$\frac{13}{15}$ & $\frac{2}{15}$
                                                  \\ & & & & & \\
$0d - 0f$  & 2 & -$\frac{3}{5}\sqrt{\frac{7}{2}}$ &
        $\frac{6}{5}\sqrt{\frac{2}{7}}$ & -$\frac{2}{15}\sqrt{\frac{2}{7}}$
                                                  &  \\ & & & & & \\
         & 4 & -$\frac{11}{15}\sqrt{\frac{2}{7}}$ &
                       $\frac{2}{15}\sqrt{\frac{2}{7}}$ &  &   \\ & & & & & \\
$0d - 1p$ &  2 & 0 & -$\frac{3}{5}$ & $\frac{2}{15}$ & \\ & & & & & \\
$0f - 0d$  & 2 & -$\frac{3}{5}\sqrt{\frac{7}{2}}$ &
        $\frac{6}{5}\sqrt{\frac{2}{7}}$ & -$\frac{2}{15}\sqrt{\frac{2}{7}}$
                                                  &  \\ & & & & & \\
         & 4 & -$\frac{11}{15}\sqrt{\frac{2}{7}}$ &
                       $\frac{2}{15}\sqrt{\frac{2}{7}}$ &  &   \\ & & & & & \\
$0f - 1p$  & 3 & $\frac{3}{5}\sqrt{\frac{2}{7}}$ &
                              -$\frac{13}{15}\sqrt{\frac{2}{7}}$
                     & $\frac{2}{15}\sqrt{\frac{2}{7}}$ & \\ & & & & & \\
$1p - 0d$  & 2 & $\frac{14}{15}$ & -$\frac{13}{15}$  & $\frac{2}{15}$ &
                                                    \\ & & & & & \\
$1p - 0f$  & 3 & $\frac{9}{5}\sqrt{\frac{2}{7}}$ &
               -$\frac{17}{15}\sqrt{\frac{2}{7}}$  &
                            $\frac{2}{15}\sqrt{\frac{2}{7}}$ &  \\ & & & & &\\
                                          \hline
\end{tabular}
\caption{Geometrical coefficients
${\zeta_{\mu}^-}(n_1l_1n_2l_2,L)$ which determine the radial
integrals ${\Braket{n_1l_1}{\theta_l^-(\rho)}{n_2l_2}}$ of Eq.
\protect (\ref{rad-int}). For details see the text.  }
 \end{center}
\end{table}
 \begin{table}
 \begin{center}
\begin{tabular}{|c|c|c|c|c|c|}  \hline\hline
 & & & & & \\
 {\em $n_1l_1 - n_2l_2$} &
 \multicolumn{1}{c|}{\em L} &
 \multicolumn{1}{c|}{\em $\mu=0$} &
 \multicolumn{1}{c|}{\em ${\mu=1}$}  &
 \multicolumn{1}{c|}{\em ${\mu=2}$}  &
 \multicolumn{1}{c|}{\em ${\mu=3}$}  \\ \hline
 & & & & & \\
$0d - 0d$ &  1 & $\frac{1}{2}$ & $\frac{4}{15}$ & -$\frac{2}{15}$  &
                                                       \\ & & & & & \\
          & 3 & $\frac{1}{15}$ & $\frac{2}{15}$ &   &
                        \\ & & & & & \\
$0f - 0f$ & 1 & $\frac{5}{6}$ & -$\frac{1}{15}$ &
                           -$\frac{26}{105}$ & $\frac{4}{105}$ \\ & & & & & \\
    & 3 & $\frac{9}{35}$ & $\frac{16}{105}$  & -$\frac{4}{105}$ &
                                                        \\ & & & & & \\
    & 5 & $\frac{2}{105}$ & $\frac{4}{105}$  &   &
                              \\ & & & & & \\
$1p - 1p$    & 1 & $\frac{1}{6}$ & $\frac{7}{15}$ &
                                  -$\frac{7}{15}$ & $\frac{2}{15}$
                                                     \\ & & & & & \\
$0d - 0f$  & 2 & $\frac{7}{6}\sqrt{\frac{2}{7}}$ &
        $\frac{4}{15}\sqrt{\frac{2}{7}}$ & -$\frac{2}{15}\sqrt{\frac{2}{7}}$
                                                  &  \\ & & & & & \\
         & 4 & $\frac{1}{5}\sqrt{\frac{2}{7}}$ &
                       $\frac{2}{15}\sqrt{\frac{2}{7}}$ &  &   \\ & & & & & \\
$0d - 1p$  & 2 & -$\frac{2}{5}$ & -$\frac{1}{5}$  & $\frac{2}{15}$ &
                                       \\ & & & & & \\
$0f - 0d$  & 2 & $\frac{1}{15}\sqrt{\frac{7}{2}}$ &
        $\frac{8}{15}\sqrt{\frac{2}{7}}$ & -$\frac{2}{15}\sqrt{\frac{2}{7}}$
                                                  &  \\ & & & & & \\
         & 4 & -$\frac{1}{15}\sqrt{\frac{2}{7}}$ &
                       $\frac{2}{15}\sqrt{\frac{2}{7}}$ &  &   \\ & & & & &\\
$0f - 1p$  & 3 & -$\frac{1}{5}\sqrt{\frac{2}{7}}$ &
                              -$\frac{7}{15}\sqrt{\frac{2}{7}}$
                                  & $\frac{2}{15}\sqrt{\frac{2}{7}}$ &
                                       \\ & & & & & \\
$1p - 0d$  & 2 & $\frac{4}{15}$ & -$\frac{1}{5}$  & $\frac{2}{15}$ &
                                       \\ & & & & & \\
$1p - 0f$  & 3 & -$\frac{1}{15}\sqrt{\frac{2}{7}}$ &
               -$\frac{1}{5}\sqrt{\frac{2}{7}}$  &
                            $\frac{2}{15}\sqrt{\frac{2}{7}}$ &
                                       \\ & & & & & \\ \hline
\end{tabular}
\caption{Geometrical coefficients $\zeta_{\mu}^+(n_1l_1n_2l_2,L)$
which determine the radial integral
$\Braket{n_1l_1}{\theta_l^+(\rho)}{n_2l_2}$ of Eq. \protect
(\ref{rad-int}). For details see the text.}
 \end{center}
\end{table}

\end{document}